\documentstyle[preprint,prl,aps,epsf]{revtex}

\begin{document}

\title{Flow at AGS Energies: A Barometer for High Density Effects?}

\author{D.~E.~Kahana$^{1}$, Y.~Pang$^{2,3}$, E.~Shuryak$^{1}$}
\address{$^{1}$Physics Department, State University of New York, \\
   	Stony Brook, NY 11791, USA\\
	$^{2}$Physics Department, Brookhaven National Laboratory\\
	Upton, NY 11973, USA\\
   	$^{3}$Physics Department, Columbia University, \\
   	New York, NY 10027, USA}  
\date{\today}  
  
\maketitle  
  
\begin{abstract}  
Preliminary data on transverse energy `flow' and event asymmetries reported
by the E877(814) collaborations are compared to ARC model calculations
for Au+Au at full AGS beam energy. ARC triple differential cross-sections for
protons and pions are presented. Proton flow is produced in ARC, with the
maximum $\langle P_x \rangle \sim 120$ MeV/c. For central events $\langle P_x
\rangle$ for the pions is near zero, consistent with experiment. Pion
$\langle P_x \rangle$ opposite to the nucleons' is evident in a peripheral
sample, however, suggesting that this pion `anti-flow' involves absorption by
`spectator' matter. `Squeeze-out' of protons in central events at
mid-rapidity is suggested by the ARC distributions.
\end{abstract}   

It is presently of great interest to know how model calculations compare with
data on flow at the AGS, especially since the excitation function of flow may
well reveal new physics, if such is occuring at high baryon density and
temperature\cite{us-flow,ES-Rischke-Gyulassy}. ARC model calculations have
previously been compared\cite{us-flow} to Bevalac data on `Sideways'
flow. Other aspects of flow, including the so-called longitudinal and radial
flows, have also been extensively studied at Bevalac energies.
Our study of Bevalac data was done at beam momenta in the range $p=$ 1--2
GeV/c, for the heavy system Au+Au, energies an order of magnitude less than
those for which the cascade was designed. Yet, very good agreement with
observed flow was obtained, using a physically plausible treatment of two body
scatterings.

Here we are concerned with recent measurements carried out at the
AGS\cite{e877-et-flow}, by the E877(814) collaboration. For details of the
experimental apparatus and other information we refer the reader
to \cite{e877-pcal-details}. For our purposes it suffices to say that
the instrument includes two highly segmented calorimeters, capable of
measuring the energy deposition at various polar and azimuthal angles,
as well as a forward spectrometer capable of particle identification.
We will discuss calorimeter data on the azimuthal asymmetry of transverse energy
as well as preliminary indications on the directed transverse momentum
$<P_x>$ measured in the spectrometer, using the reaction plane determined
from the calorimeters. 

The two calorimeters are TCAL and PCAL (target and participant calorimeters),
with pseudorapidity coverages of $-0.5 < \eta < 0.8$ and $0.83 < \eta < 4.5$,
respectively. Only PCAL is used in constructing the transverse energy
distributions. The dispersion in the event plane determined from the PCAL is
on the order of $60^o$ (FWHM) from experiment\cite{PBM-private}, and theoretical
simulation using ARC events yields a similar result. Therefore relatively
large corrections to the raw experimental numbers are needed when extracting
the triple differential cross-sections. Detector acceptances produce
additional uncertainty. Nevertheless, the preliminary results clearly
demonstrate the existence of proton flow at AGS, and give some hint as to the
quantitative amount. For pions one observes, interestingly, no flow in a
central sample\cite{Wessels-private}.

Centrality cuts are placed on data and on ARC events using the total PCAL
$E_T$. Central events produce more transverse energy than peripheral. For ARC
we calculated PCAL $E_T$ using just the angular acceptance, not accounting
for any efficiency corrections. The best procedure, of course, would be to
pass ARC events through the experimental simulation of the PCAL response.
However, the agreement in the $E_T$ spectrum calculated directly is very close,
certainly at better than $10\%$ level for all but the most peripheral events.
For the purpose of making centrality cuts on the flow measurements
we have assumed that E877 $E_T$ and (raw) ARC $E_T$ are the same.

The azimuthal transverse energy distribution from E877 has been
analysed \cite{e877-et-flow}, using an expansion in spherical
harmonics. Moments $v_n$ of the distribution of $E_T(\phi)$ are
computed, by summing over the various cells $\lbrace i\rbrace$ composing the
PCAL, which is nearly symmetric in $\phi$:
$$
v_n = \frac{\sum_i E^i_T \cos(n\phi_i)}{\sum_i E^i_T}
$$
Moments up to $v_4$ have been considered.
Including the corresponding sine transforms, one can define two-dimensional
vectors $v_n$. The direction of $v_1$ for example, is an estimate of the
direction of the reaction plane in a particular event. In principle then, the
distributions of the relative phases of the $v_n$ are meaningful too.

A more transparent procedure might have been to fit an ellipse to the
distribution $E_T(\phi)$. The center, the major and minor axes,
and the orientation of the ellipse would provide a small set of physically
meaningful parameters by which to characterise the distributions.
The finite detector segmentation in $\phi$ limits the highest
frequency $v_n$ containing significant information. In addition,
fluctuations are large on an event by event basis, and high frequency
noise is present, further obscuring the meaning of higher moments. We
consider it unlikely that much real information is present beyond
$v_1$ and $v_2$, and it is hard to imagine what physics could lead to
such azimuthal distributions. In any case, only $v_1$ shows any
significant systematic effects (dependence on centrality for example)
in the present data.

We have calculated $v_1$ and $v_2$ from ARC events, simulating the
segmentation of PCAL using sixteen bins in $\phi$. The pseudorapidity
interval considered is $1.85 < \eta < 4.5$.
 
\begin{figure}
\vbox{\hbox to\hsize{\hfil
 \epsfxsize=6.1truein\epsffile[85 207 581 549]{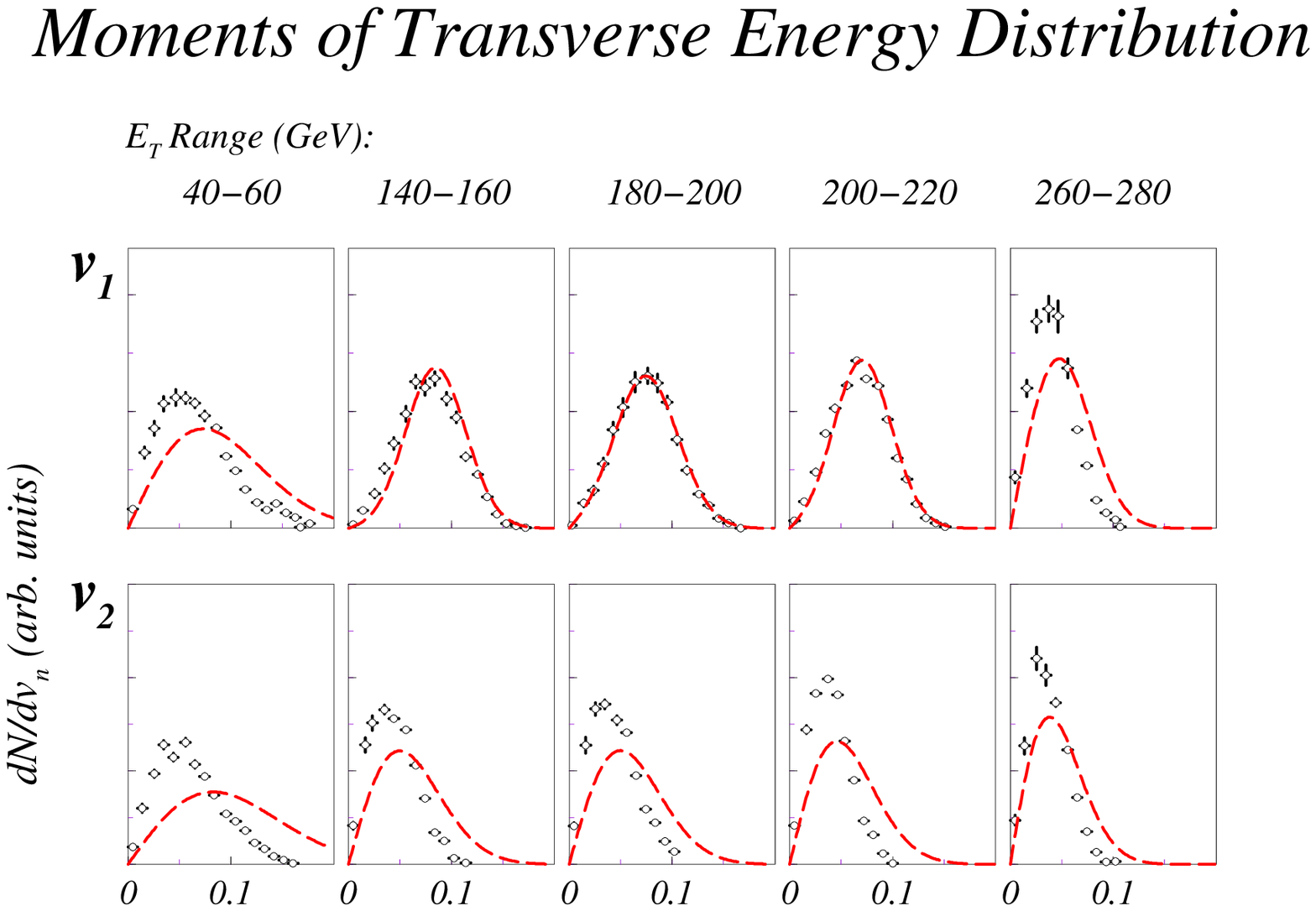}
 \hfil}}
\caption[] {Moments of the distribution $E_t(\phi)$.}
\label{Fig:moments}
\end{figure}

A comparison of the ARC moments with data is shown in Fig(\ref{Fig:moments}).
One can see clearly that the distributions of $v_n$ calculated from ARC agree
quite well with the experimental ones, both in shape and in dependence on
centrality.  We plot only the magnitudes of $v_n$, and the Jacobian factor
$(1/v_n)$ is not included. The characters of the two dimensional
distributions for $v_1$ and $v_2$ differ: the $v_2$ distributions have a
maximum at $\vert v_2 \vert = 0$ for all centralities. The distributions for
$v_1$ are also peaked at zero for the most peripheral and most central event
samples. But for semi-central events $v_1$ has a maximum {\it away} from
zero. This is just the expected systematics of flow with centrality, the
position of the maximum in $v_1$ being roughly related to the size of the
transverse flow.

Plots of the triple differential cross sections from ARC, for an event sample
corresponding to the E877 central cut (10\% of PCAL $E_t$) are in
Fig(\ref{Fig:montage_c}). For this cut, no flow is visible in the pion
spectra. The pion distributions are symmetric around the beam axis, seemingly
consistent with experiment\cite{Wessels-private}. The proton spectra
shift progressively towards positive $\langle P_x \rangle$ with increasing
rapidity. Additionally, at mid-rapidity the distribution is
distorted, appearing roughly elliptical, with the major axis
oriented along the $\langle P_y \rangle$ direction. This anisotropy vanishes
near beam rapidity, where one sees only the shift toward positive
$\langle P_x \rangle$. This would appear to be the so-called
`squeeze out', in which protons are ejected symmetrically out of the
reaction plane near mid-rapidity. `Squeeze out' has not been not observed as
yet at the AGS (it would be difficult at best to detect by analysing the
$v_n$), but we note that quite a good definition of the
reaction plane is probably necessary to discern it. Here, we are
presenting essentially raw results from ARC. Folding these results with
detector acceptances could possibly have a large effect.

For the peripheral event sample Fig (\ref{Fig:montage_p}) the triple
differential cross-sections are quite different. Most strikingly, the pion
distributions show a clear shift toward {\it negative} $\langle P_x \rangle$,
{\it i.~e.}~opposite to the protons.  This is the so-called `anti-flow', and that it
appears in the peripheral sample but is absent in the central one suggests
that the effect is in large part due to shadowing by spectator nucleons. We note
that E866 has detected an asymmetry in the pion distributions for
semi-peripheral events\cite{e802-asymmetries}.

Our previous study of heavy ion data at BEVALAC energies \cite{us-flow} using
ARC has shown that a good description of the observed flow can be
obtained across a range of energies and projectiles. Central to
achieving this agreement at low energy was a correct treatment of $pp$
scattering, specifically the collisions leading to two-body final states. 
Kinetic pressure in the cascade was found to produce on
the order of $90$\% of the observed flow in the heaviest systems, once
angular momentum conservation was properly taken into account by preserving
the reaction plane in such collisions. If the reaction plane were
randomized instead, about $70$\% of the flow would be obtained. Only an
additional $10$\% difference in the flow is ascribed to a potential-like
effect in two-body scattering\cite{us-flow}.  Proton rapidity and
transverse mass spectra were also found to be in agreement with data and
to be nearly independent of the choice of orbits and the randomisation
of the reaction plane. The conclusion is that at these low energies little
potential or mean field enters: the pure cascade accounts for the flow.

Because the $NN$ interaction is mostly repulsive at the higher LBL energies,
the classical cascade can represent scattering well using dominantly
repulsive outgoing paths for two-body final states. In actual potential
modeling the character of the classical orbits will vary with impact
parameter and energy. One expects repulsive orbits for small
impact parameters probing the inner core, and attractive orbits passing
through the outer regions. At sufficiently high energy the description in
terms of real potential scattering breaks down, and pion production
becomes the dominant process. A reasonable guess is that somewhere above
the threshhold for two pion production it is not sensible anymore to
talk about potential models. In ARC, production of up to two pions
$pp\rightarrow n_\pi\pi$ proceeds entirely through excitation of the
$\Delta \Delta$ intermediate state. Hence this scattering can still be
described in terms of two baryon states. For first collisions at AGS energies
the average pion number has risen to $3.8$ per inelastic collision,
the two body optical theorem is violated and no description in terms
of potential scattering is possible. The potential, if any, is clearly
dominated by its absorptive, imaginary part, and the scattering has become
diffractive in nature. What then can one say about the nature of the
classical orbits at such high energies? Diffraction from a black disk
occurs in the absence of either repulsion or attraction in the potential,
therefore the best guess is probably that the orbits should be equally
often repulsive or attractive.

In constructing a cascade to be used at all energies between BEVALAC and AGS,
one must interpolate between the mostly repulsive $NN$ scattering occuring at
low energies and the diffractive $NN$ scattering taking place at higher
energies. Heavy ion collisions at the AGS involve individual $NN$ scatterings
taking place at all energies equal to or below the first collision
energy. The simplest procedure which builds in the essential physics is to
use repulsive orbits below some energy, $E_{min}$, a 50/50 mixture of
repulsive and attractive orbits above a higher energy, $E_{max}$, and to
interpolate smoothly between these limits. In practice we take
$E_{min}\sim 300$ MeV and $E_{max}\sim 500$ MeV corresponding crudely to
the effective $2\pi$ and $3\pi$ thresholds. We can distinguish this approach
quite sharply from one involving mean fields: what we propose is not at all
a medium dependent effect, rather it is an energy dependent treatment of the
{\it free space} two body scattering. This is justified, we think, by the
nature of the $NN$ interaction.

\begin{figure}
\vbox{\hbox to\hsize{\hfil
 \epsfxsize=6.1truein\epsffile[30 95 542 714]{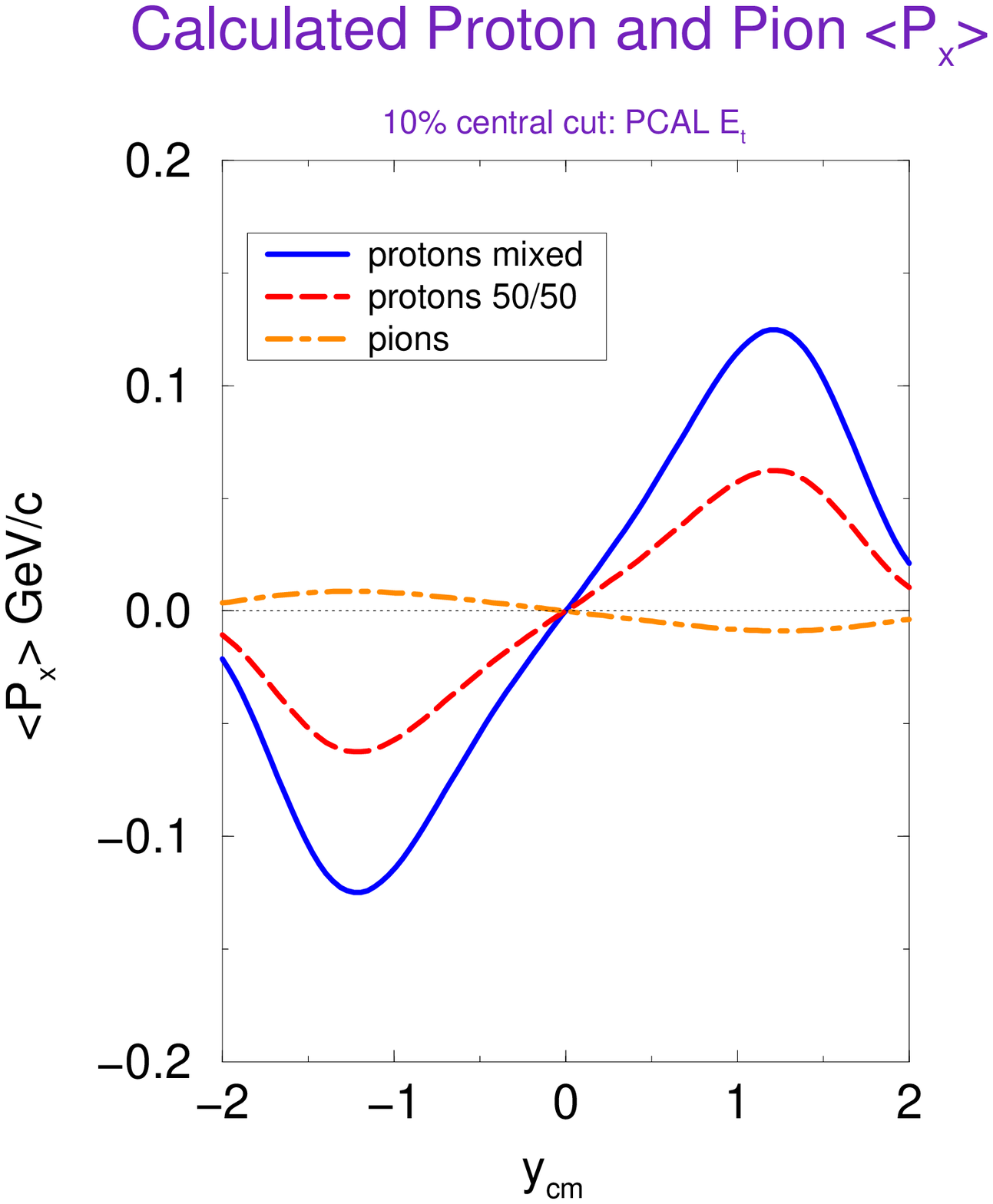}
 \hfil}}
\caption[]{$\langle P_x \rangle$ for protons and pions. Central (10\%)
	cut based on PCAL $E_t$. Energy dependent mixture of
	repulsive scattering against 50/50 mixture
	at all energies.}
\label{Fig:px_orbits}
\end{figure}

In Fig(\ref{Fig:px_orbits}) we display the effect on $\langle P_x\rangle$ of
choosing an energy dependent mixture of repulsive and attractive orbits
versus the 50/50 mixture. Choosing purely repulsive orbits at BEVALAC
energies produced a rather small ($<10$\%) increase in flow discussed
above. At AGS energies, however, the sensitivity to the classical orbits
can be much larger, resulting in perhaps as much as $60\%$ change in $\langle
P_x\rangle$ for a central cut. Indeed, the `potential' one would need to
turn all the orbits outward for $p=10.8$ GeV/c would have to be very large
indeed and completely unrealistic. If all the orbits are made attractive,
interestingly, the flow is reduced to near zero.

Summarising, present analysis of the very interesting E877 experiments on
azimuthal asymmetry, combined with previous success at the BEVALAC, lends
confidence to the cascade description of sideways flow. Certainly, the ARC
calculations are in substantial agreement Fig.(\ref{Fig:moments}) with the
higher energy data. The absolute flow magnitude is, however, more sensitive
to the potential contribution at $11.6$ GeV/c than at $2$ GeV/c.  The method
of interpolation between real potential and diffractive two body scattering
is straightforward, but its full validation awaits the detailed E877 flow
results and triple differential cross-sections. Nevertheless, it is
interesting to speculate on the predictions of this model at intermediate
beam momenta, $2-8$ GeV/c, relative to ongoing experiments with the EOS TPC
\cite{eos}, now installed at the AGS. At these, lower, AGS energies for
$Au+Au$ one may reasonably expect to form a largely equilibrated compound
system, still at high baryon density. So, the likelihood of nucleating
appreciable amounts of quark-gluon plasma\cite{kapusta} should be
optimized. The signature would be a considerable reduction in flow following
from a reduction in kinetic pressure. The pioneering BEVALAC, E877 and E866
experiments provide needed guidance, fixing the flow at both low energy and
at the full AGS $Au$ energy, and setting the stage for perhaps dramatic
results in between.

The authors thank S.H.Kahana for useful discussions. This work has been
supported by DOE grants No. DE-FG02-93~ER40768, DE-AC02-76~CH00016, and
DE-FG02-92~ER40699.

\begin{figure}
\vbox{\hbox to\hsize{\hfil
 \epsfxsize=6.0truein\epsffile[65 190 540 620]{1fig2.ps}
 \hfil}}
\caption[]{Proton and pion transverse momentum distributions relative
	to the event plane (x-direction) for 10\% central events. Pion
	distributions are symmetric at all rapidities, indicating no
	pion anti-flow at this centrality. This seems to be consistent
	with the current experimental picture. Proton distributions show
	a clear shift towards positive $\langle P_x\rangle$. A spectator
	component is present in the proton distribution for the most
	forward rapidity. The deformation of the proton distributions
	along $\langle P_y \rangle$ indicates proton squeezout occurs at
	mid-rapidity.}
\label{Fig:montage_c}
\end{figure}
\clearpage
\begin{figure}
\vbox{\hbox to\hsize{\hfil
 \epsfxsize=6.1truein\epsffile[65 190 540 620]{1fig3.ps}
 \hfil}}
\caption[]{Peripheral pion and proton momentum distributions ($b>3$ fm).
	Proton distributions appear symmetric at all rapidities:
	$\langle P_x \rangle$ is small for peripheral samples.
	A large spectator component is present at the most forward rapidity.
	Pion distributions show a distinct shift to negative $\langle P_x
	\rangle$: that this pion `anti-flow' appears only in the peripheral
	sample indicates it is largely due to shadowing.}
\label{Fig:montage_p}
\end{figure}
\clearpage
%
%
\clearpage
\end{document}